\documentclass[12pt]{iopart}
\usepackage{graphicx}
\eqnobysec

\newcommand{\abs}[1]{\vert#1\vert}
\newcommand{\ave}[1]{\overline{#1}}
\newcommand{\bigabs}[1]{\left\vert#1\right\vert}
\newcommand{\blanc}{{\vphantom{\ds M}}}
\newcommand{\cm}{{\rm cm}}
\newcommand{\dd}{{\rm d}}
\newcommand{\deuxlignes}[2]{$\ds{\matrix{\hbox{#1}\cr\hbox{#2}}}$}
\newcommand{\ds}{\displaystyle}
\renewcommand{\e}{{\rm e}}
\newcommand{\eps}{\varepsilon}
\newcommand{\frad}[2]{\ds{\frac{\ds#1}{\ds#2}}}
\newcommand{\ii}{{\rm i}}
\newcommand{\mean}[1]{\langle#1\rangle}

\newcommand{\n}{{\bf n}}
\newcommand{\s}{\sigma}

\begin{document}

\title
[Dynamics in low-dimensional disordered systems with extended states]
{Dynamics of a quantum particle in low-dimensional disordered systems
with extended states}

\author{P L Krapivsky$^{1,2}$ and J M Luck$^2$}

\address{$^1$ Department of Physics, Boston University, Boston, MA 02215, USA}

\address{$^2$ Institut de Physique Th\'eorique, IPhT, CEA Saclay
and URA 2306, CNRS, 91191 Gif-sur-Yvette cedex, France}

\begin{abstract}
We investigate the dynamics of a quantum particle in disordered tight-binding models
in one and two dimensions which are exceptions to the common wisdom
on Anderson localization,
in the sense that the localization length diverges at some special energies.
We provide a consistent picture for two well-known one-dimensional examples:
the chain with off-diagonal disorder and the random-dimer model.
In both cases the quantum motion exhibits a peculiar kind
of anomalous diffusion which can be referred to as bi-fractality.
The disorder-averaged density profile of the particle becomes critical
in the long-time regime.
The $q$-th moment of the position of the particle diverges with time
whenever~$q$ exceeds some $q_0$.
We obtain $q_0=2$ for off-diagonal disorder on the chain
(and conjecturally on two-dimensional bipartite lattices as well).
For the random-dimer model, our result $q_0=1/2$ corroborates known rigorous results.
\end{abstract}


\eads{\mailto{pkrapivsky@gmail.com},\mailto{jean-marc.luck@cea.fr}}

\maketitle

\section{Introduction}

Over fifty years after Anderson's seminal paper on localization~\cite{fifty},
the common wisdom is that in one dimension any disorder
drives the phenomenon of Anderson localization.
Even when the strength of disorder is arbitrarily small,
all the eigenstates of the Hamiltonian get exponentially localized.
The same picture applies to a large extent to the two-dimensional case as well.

In a number of low-dimensional disordered systems the localization length $\xi(E)$
is however known to diverge at special energies.
This phenomenon is usually the manifestation of some symmetry.
A well-known example is that of tight-binding models with off-diagonal disorder,
both in one dimension~\cite{D,HJ,TC,FL,ER,E}
and on higher-dimensional bipartite lattices~\cite{G,ITA,H,M}.
In this context the localization length becomes infinite at the center of the band
($E\to0$).
This phenomenon is due to particle-hole symmetry ($E\leftrightarrow-E$).
In the one-dimensional situation, it was already investigated in the pioneering work
by Dyson~\cite{D} on disordered harmonic chains,
while the interest in two-dimensional models has been revived
by a connection with the integer Quantum Hall Effect~\cite{W,F}.

It is natural to wonder what are the dynamical consequences of the divergence
of the localization length at the band center in this class of models.
The usual setting consists in looking at the spreading of the wave-packet
of a particle initially launched at the origin.
A variety of rigorous general results,
following the pioneering work by Guarneri~\cite{Gua},
relate the growth of the moments
of the position of the particle to dimensions of the spectral measure.
The latter results however have very little predictive power in the present situation.

The goal of this work is to obtain quantitative, albeit non-rigorous dynamical results,
which can be checked against numerical data.
Our main finding will be that the quantum motion exhibits a peculiar kind
of anomalous diffusion which can be referred to as bi-fractality~\cite{L}.
The disorder-averaged density profile $\rho_n(t)$ of the particle at time $t$
becomes critical in the long-time regime,
in the sense that it falls off as a power of distance,
as $\rho_n=\lim_{t\to\infty}\rho_n(t)\sim1/\abs{n}^{q_0+1}$.
As a consequence, the moments
$M_q(t)=\ave{\mean{\abs{n}^q}_t}=\sum_n\abs{n}^q\,\rho_n(t)$
of the position of the particle
diverge with time $t$ whenever $q$ exceeds~$q_0$,
while they have finite asymptotic values for $q<q_0$.
For the chain with off-diagonal disorder, investigated in Section~2, we obtain $q_0=2$.

Another example of interest is the random-dimer model~\cite{PP,B,FH,Sveta}.
This is a one-dimensional tight-binding model with diagonal binary disorder,
so that the site energies $V_n=\pm V$ assume the same values on dimers,
i.e., pairs of consecutive sites.
The localization length is known to diverge at the special energies $E=\pm V$ when $V\le1$,
in units where the hopping amplitude is one.
We demonstrate in Section~3 that the dynamics of the random-dimer model
fits within exactly the same framework, with $q_0=1/2$.
These findings confirm earlier heuristic results~\cite{PP},
namely $M_2(t)\sim t^{3/2}$ in the generic situation ($V<1$),
while $M_2(t)\sim t$ in the critical situation ($V=1$).
In the first case, our predictions are in agreement with
recent rigorous results~\cite{Sveta}.

Section~4 is devoted to a summary and discussion of our findings.
We also argue why the same framework is expected to describe the quantum dynamics
on two-dimensional bipartite lattices with off-diagonal disorder.

\section{The chain with off-diagonal disorder}

In this section we investigate the time-dependent
discrete Schr\"odinger (tight-binding) equation
on the one-dimensional chain with off-diagonal disorder:
\begin{equation}
\label{SE1D}
\ii\,\dot{\psi}_n=V_n\psi_{n+1}+V_{n-1}\psi_{n-1}.
\end{equation}
The hopping amplitudes $V_n$ between sites $n$ and $n+1$
are modelled as i.i.d.~(independent and identically distributed) random variables.
We set\footnote{Our definition of $\s^2$ differs from that used in~\cite{ER} by a factor 4.}
\begin{equation}
\label{epsdef}
V_n=\e^{\eps_n},\qquad\ave{\eps_n\vphantom{h}}=0,\qquad\ave{\eps_n^2}=\s^2.
\end{equation}

\subsection{Spectral properties revisited}

The time-independent (spectral) problem,
defined by the eigenvalue equation
\begin{equation}
\label{SES}
E\psi_n=V_n\psi_{n+1}+V_{n-1}\psi_{n-1},
\end{equation}
has been the subject of much activity since the pioneering work of Dyson~\cite{D}.
The tight-binding equation~(\ref{SES}) has the peculiar feature
that the even and odd sub-lattices decouple at zero energy.
Hence the general solution at $E=0$ can be written down explicitly:
\begin{equation}
\label{zeroE}
\psi_{2p}=(-1)^p\,\e^{B_{2p}}\psi_0,\qquad
\psi_{2p+1}=(-1)^p\,\e^{B_1-B_{2p+1}}\psi_1,
\end{equation}
where $B_n$ is the following random walk (discrete Brownian motion)
associated with the disorder:
\begin{equation}
\label{Bdef}
B_n=\sum_{m=0}^{n-1}(-1)^m\eps_m.
\end{equation}
The wavefunction~(\ref{zeroE}) exhibits unusual scaling properties~\cite{FL}.
We have $\ave{B_n^2}=\s^2 n$,
and so the zero-energy wavefunction typically grows sub-exponentially
as $\ln\abs{\psi_n}\sim\s n^{1/2}$.
In other words, the effective localization length grows with distance as
\begin{equation}
\label{xieff}
\xi(n)\sim\frac{n^{1/2}}{\s},
\end{equation}
so that in the thermodynamic limit, the localization length is infinite at zero energy.

This peculiarity affects the behavior of various quantities near the band center.
The Lyapunov exponent $\gamma(E)$ at energy $E$
and the integrated density of states $H(E)$ up to energy $E$
exhibit the following Dyson singularities~\cite{D,TC,ER} as $E\to 0$:
\begin{equation}
\label{Dyson}
\gamma(E)\approx\frac{\s^2}{-\ln\abs{E}},\qquad
\bigabs{H(E)-\frac12}\approx\frac{\s^2}{2(\ln\abs{E})^2}\,.
\end{equation}
The first result implies that the localization length $\xi(E)=1/\gamma(E)$
exhibits a logarithmic divergence as energy $E$ approaches the band center,
whereas the second one shows that states pile up near the band center,
as the density of states strongly diverges as $E\to 0$~as
\begin{equation}
\label{rhoDyson}
\rho(E)\approx\frac{\s^2}{\abs{E}(-\ln\abs{E})^3}\,.
\end{equation}
This expression is at the verge of being non integrable.

It will prove useful to first consider the finite-size scaling consequences
of the Dyson singularity.
On a finite chain made of an even number $N$ of sites,
with periodic boundary conditions,
the spectrum manifests an exact particle-hole symmetry~\cite{ITA,B+}:
if $\psi_n$ is an eigenvector for the energy eigenvalue $E$,
then $(-1)^n\psi_n$ is an eigenvector for the opposite energy~$(-E)$.
Furthermore
the expressions~(\ref{zeroE}) show that $E=0$ can be an eigenvalue
only if~$N$ is a multiple of 4.
This eigenvalue is then twofold degenerate.
The condition on the disorder for this to occur is $B_N=0$,
i.e., $V_1V_3\dots V_{N-1}=V_2V_4\dots V_N$.

We denote by $E_1$ the energy gap of a finite system,
defined as the smallest positive energy eigenvalue.
The form~(\ref{Dyson}) of the integrated density of states suggests
to set $X_1=-\ln E_1$.
Typical values of the latter quantity can be expected to be
such that $H(E_1)-1/2\approx\s^2/(2X_1^2)\sim1/N$.
We thus obtain the finite-size scaling law
\begin{equation}
\label{Xfss}
X_1=-\ln E_1\approx\s\,N^{1/2}\,Y_1,
\end{equation}
where $Y_1$ is a fluctuating quantity of order unity.
The eigenstates at energy $\pm E_1$ are expected to be the most extended ones.
The Dyson singularity~(\ref{Dyson})
implies that their localization length scales as $\xi_1\sim X_1/\s^2\sim N^{1/2}/\s$.
This result is in agreement with the estimate~(\ref{xieff})
of the effective distance-dependent localization length $\xi(n)$ at zero energy,
taken at the maximal distance $n\sim N$.

\begin{figure}[!ht]
\begin{center}
\includegraphics[angle=-90,width=.45\linewidth]{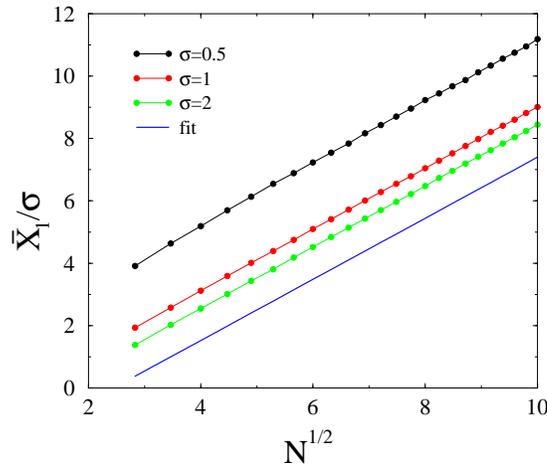}
\caption{\label{fx}
Plot of $\ave{X_1}/\s$ against $N^{1/2}$,
for three values of the disorder strength~$\s$.
The blue line is the result of a common least-square fit to all datasets.}
\end{center}
\end{figure}

Figure~\ref{fx} shows a plot of $\ave{X_1}/\s$ against $N^{1/2}$.
The energy gap $E_1$ is obtained by
a numerical diagonalization of the Hamiltonian matrix on finite chains
of $N$ sites, with periodic boundary conditions.
Only values of $N$ multiples of 4 have been used.
[Data for $N\equiv 2~(\hbox{mod}~4)$ exhibit
stronger corrections to the asymptotic linear behavior.
This phenomenon can be related to the preceding observation that $E=0$ can only be
an exact eigenvalue if $N$ is a multiple of 4.]
In numerical simulations hereinafter,
the $\eps_n$ are drawn uniformly
in the interval $-W/2<\eps_n<W/2$, with $W^2=12\s^2$.
We shall report results for
three values of the disorder strength: $\s=0.5$, 1, and 2.
Data are averaged over $10^5$ samples for each value of $N$ and $\s$.
The observed linear behavior corroborates the scaling law~(\ref{Xfss}).
The slope of a common least-square fit to all datasets (blue line)
yields $\ave{Y_1}\approx0.98$.

\begin{figure}[!ht]
\begin{center}
\includegraphics[angle=-90,width=.45\linewidth]{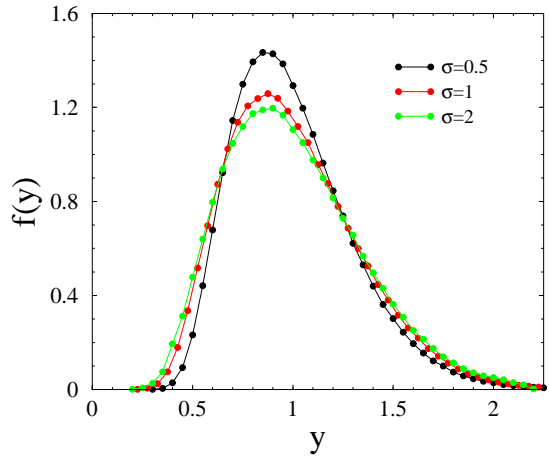}
\caption{\label{fxhisto}
Histogram plot of the distribution of $y=Y_1/\ave{Y_1}=X_1/\ave{X_1}$ for $N=100$.}
\end{center}
\end{figure}

Figure~\ref{fxhisto} shows a histogram plot of the distribution
of the reduced variable $y=Y_1/\ave{Y_1}=X_1/\ave{X_1}$
for the largest system considered, i.e., $N=100$.
This distribution is observed to fall off very fast,
both at small and large values of the variable,
and to have a rather weak dependence on $\s$.

\subsection{Dynamical properties}

Let us now return to the quantum dynamics of the particle
described by the time-dependent equation~(\ref{SE1D}).
We assume that the particle starts at the origin:
\begin{equation}
\label{t=0}
\psi_n(0)=\delta_{n,0}\,.
\end{equation}
Our basic observables will be the moments of the position of the particle at time $t$:
\begin{equation}
\label{Mqdef}
M_q(t)=\ave{\mean{\abs{n}^q}_t}=\sum_n\abs{n}^q\,\rho_n(t),
\end{equation}
where
\begin{equation}
\label{rhodef}
\rho_n(t)=\ave{\abs{\psi_n(t)}^2}
\end{equation}
is the disorder-averaged density profile at time $t$,
whereas $q$ is an arbitrary positive index.
Hereinafter,
the brackets $\mean{\cdots}_t$ denote an average over the quantum state at time~$t$
described by the wavefunction $\psi_n(t)$,
whereas the bar $\ave{\cdots\vphantom{h}}$ denotes an ensemble average over the disorder,
i.e., over the distribution of the random hopping amplitudes~$\{V_n\}$.

On a finite sample made of $N$ sites, the moments
\begin{equation}
\label{sum0}
\mean{\abs{n}^q}_t=\sum_n\abs{n}^q\,\abs{\psi_n(t)}^2
\end{equation}
can be expressed in terms of the $N$ energy eigenvalues $E^\alpha$
and of the corresponding (real and normalized) eigenvectors $\psi_n^\alpha$.
Expanding the initial condition as
\begin{equation}
\label{psi0}
\psi_n(0)=\delta_{n,0}=\sum_{\alpha}\psi_n^\alpha\,\psi_0^\alpha,
\end{equation}
we find the wavefunction at arbitrary time $t\geq 0$:
\begin{equation}
\label{psit}
\psi_n(t)=\sum_\alpha\e^{-\ii E^\alpha t}\psi_n^\alpha\,\psi_0^\alpha.
\end{equation}
Combining~(\ref{sum0}) and~(\ref{psit}), we arrive at
\begin{equation}
\label{sum1}
\mean{\abs{n}^q}_t=\sum_{\alpha,\beta}\e^{-\ii(E^\alpha-E^\beta)t}
\psi_0^\alpha\,\psi_0^\beta\sum_{n}\abs{n}^q\,\psi_n^\alpha\,\psi_n^\beta.
\end{equation}

The long-time behavior of the above exact expression
can be analyzed in the following heuristic way,
along the line of thought of~\cite{L}.
First, neglecting interference terms between different quantum states, we obtain
\begin{equation}
\label{sum2}
\mean{\abs{n}^q}_t\approx\sum_\alpha
(\psi_0^\alpha)^2\sum_{n}\abs{n}^q(\psi_n^\alpha)^2.
\end{equation}
Second, we model the eigenstates by considering that the probability density
$(\psi_n^\alpha)^2$ is roughly uniform over the range $\abs{n-n_\cm^\alpha}\le\xi^\alpha$,
where we have introduced the center-of-mass co-ordinate $n_\cm^\alpha$
and the localization length $\xi^\alpha$ of eigenstate $\alpha$, defined as
\begin{equation}
\label{nxi}
n_\cm^\alpha=\sum_n n(\psi_n^\alpha)^2,\qquad
(\xi^\alpha)^2=\sum_n(n-n_\cm^\alpha)^2(\psi_n^\alpha)^2.
\end{equation}
This picture implies that the eigenstates which contribute to the sum~(\ref{sum2})
are those such that $\abs{n_\cm^\alpha}\le\xi^\alpha$.
Putting everything together,
we are left with the following asymptotic long-time estimate
for the disorder-averaged moments:
\begin{equation}
\label{Malpha}
M_q=\lim_{t\to\infty}M_q(t)\sim\frac1N\sum_\alpha\ave{(\xi^\alpha)^q}\,.
\end{equation}
Finally, for a large system,
the sum over discrete states can be replaced by an integral over the spectrum:
\begin{equation}
\label{Mt}
M_q\sim\int\xi(E)^q\,\rho(E)\,\dd E.
\end{equation}

In the customary situation of the Anderson model with diagonal disorder,
the moments of the position are thus predicted to saturate to finite limiting values
of order $M_q\sim\xi_0^q$,
where for definiteness $\xi_0$ is the localization length at the band center.
More quantitative predictions can be worked out,
including universal pre-factors in the weak-disorder regime~\cite{L}.

The present situation of the Anderson model with off-diagonal disorder
is radically different.
Indeed, the Dyson singularities~(\ref{Dyson}) which affect
the localization length and the density of states near the band center
cause the integral in~(\ref{Mt}) to diverges for sufficiently large $q$.
More precisely, taking $X=-\ln\abs{E}\gg1$ as independent variable,
the expressions~(\ref{Dyson}) yield $\xi(E)\approx X/\s^2$
and $\rho(E)\,\dd E=\dd H(E)\approx\s^2\,\dd X/X^3$, and so~(\ref{Mt}) reads
\begin{equation}
\label{MDyson}
M_q=\ave{\mean{\abs{n}^q}}
\sim\s^2\int\left(\frac{X}{\s^2}\right)^q\,\frac{\dd X}{X^3}\,.
\end{equation}
The integral is divergent for $q>2$.

Let us first consider the stationary regime of very long times,
and focus onto the integrand in~(\ref{MDyson}),
not paying attention to the integration bounds for the time being.
It is natural to identify the integration variable $X$
with the position $\abs{n}$ of the particle, according to $\abs{n}=X/\s^2$.
We can thus read off from~(\ref{MDyson})
that the asymptotic disorder-averaged density profile
reached by the particle in the long-time regime,
\begin{equation}
\label{asydef}
\rho_n=\lim_{t\to\infty}\rho_n(t),
\end{equation}
falls off as the power law
\begin{equation}
\label{rho}
\rho_n\sim\frac{1}{\s^2\,\abs{n}^3}\,.
\end{equation}

Let us move to the late stages of the dynamics of the quantum particle.
As the latter starts from the origin,
its wavefunction spreads progressively over larger and larger distances.
For a large but finite time $t$, the power-law density profile~(\ref{rho})
is therefore expected to be cutoff at some scale $n_*(t)$.
The growth law of the crossover scale $n_*(t)$
can be estimated as follows.
First, as a consequence of the uncertainty principle,
the finiteness of the observation time $t$
induces a finite energy resolution $E_*(t)\sim1/t$.
The largest observable localization length at time $t$
therefore scales as the localization length at energy $E_*(t)$
near the band center.
Hence it diverges as $\xi_*(t)\sim-(\ln E_*(t))/\s^2\sim(\ln t)/\s^2$.
Then, the cutoff scale $n_*(t)$ can be estimated by equating
the above $\xi_*(t)$ to the effective localization length $\xi(n)$
of~(\ref{xieff}).
We thus obtain
\begin{equation}
\label{nstar}
n_*(t)\sim\frac{(\ln t)^2}{\s^2}\,.
\end{equation}

The crossover scale $n_*(t)$ provides the appropriate cutoff
to the integral in~(\ref{MDyson}).
The disorder-averaged moments $M_q(t)$ of the position
can now be evaluated by means of the estimates~(\ref{rho}) and~(\ref{nstar}).
We are thus left with the prediction that the moments grow with time
on a logarithmic scale for $q\ge2$,
whereas they saturate to finite values for $q<2$:
\begin{equation}
\label{Mq}
M_q(t)\sim
\left\{\matrix{
\frad{(\ln t)^{2(q-2)}}{\s^{2(q-1)}}\hfill&(q>2),\cr
\frad{1^\blanc}{\s^2_\blanc}\ln\frac{(\ln t)^2}{\s^2}\hfill\quad&(q=2),\cr
\hbox{finite}\hfill&(q<2).
}\right.
\end{equation}
We have in particular
\begin{equation}
\label{M34}
M_3(t)\approx a_3\frac{(\ln t)^2}{\s^4},\qquad
M_4(t)\approx a_4\frac{(\ln t)^4}{\s^6}.
\end{equation}

\begin{figure}[!ht]
\begin{center}
\includegraphics[angle=-90,width=.45\linewidth]{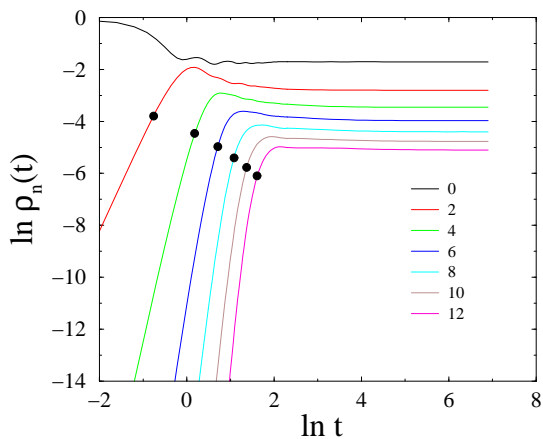}
\caption{\label{fprofil}
Log-log plot of the disorder-averaged density $\rho_n(t)$
against time $t$, for $\s=0.5$ and various even distances $n$.
Symbols: crossover times $t_*(n)$.}
\end{center}
\end{figure}

The above analysis has been tested against numerical data
generated by directly simulating the time-dependent tight-binding equation~(\ref{SE1D}).
Figure~\ref{fprofil} shows a log-log plot of the disorder-averaged density $\rho_n(t)$
against time $t$, for $\s=0.5$ and even distances~$n$ ranging from 0 to 12.
The convergence to asymptotic densities~$\rho_n$
is observed to be slower and slower for larger distances.
For each distance $n$, a black symbol marks the crossover time $t_*(n)$,
defined operationally as the time where $\rho_n(t)=\rho_n/\e$
($\e\approx2.718281$).
The oscillations which can be seen in $\rho_0(t)$, and to some extent in~$\rho_2(t)$,
are reminiscent of those which are present in the absence of disorder,
where the time-dependent wavefunction reads $\psi_n(t)=(-\ii)^nJ_n(2t)$,
with $J_n$ being the Bessel functions.
We mention for further reference that the latter wavefunction
exhibits ballistic peaks around the positions $n=\pm 2t$,
whose width and height respectively scale as $t^{1/3}$ and~$t^{-1/3}$.
More precisely, setting $\abs{n}=2t+t^{1/3}z$, one finds~\cite{L}
\begin{equation}
\label{psiairy}
\ii^n\psi_n(t)\approx t^{-1/3}{\rm Ai}(z),
\end{equation}
with ${\rm Ai}(z)$ being the Airy function.

\begin{figure}[!ht]
\begin{center}
\includegraphics[angle=-90,width=.45\linewidth]{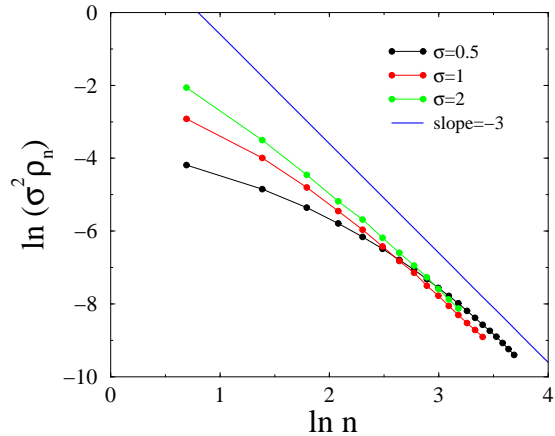}
\caption{\label{fp}
Log-log plot of $\s^2$ times the asymptotic densities $\rho_n$,
against distance $n$.
The blue line with slope $(-3)$ illustrates the theoretical prediction~(\ref{rho}).}
\end{center}
\end{figure}

Figure~\ref{fp} shows a log-log plot of~$\s^2$ times the asymptotic densities $\rho_n$,
against distance $n$.
The data corroborate the prediction~(\ref{rho}),
shown by the blue line with slope $(-3)$.
The amplitude of the $1/n^3$ law may have some weak residual dependence
on $\s$, besides the main scaling in $1/\s^2$.
The corrections to the asymptotic behavior are observed
to be larger at weaker disorder.

\begin{figure}[!ht]
\begin{center}
\includegraphics[angle=-90,width=.45\linewidth]{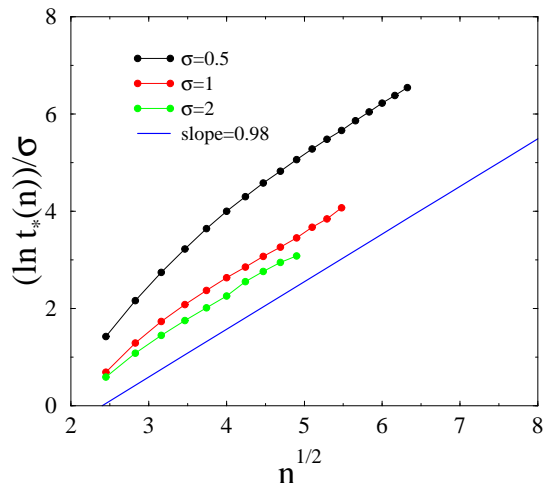}
\caption{\label{ft}
Plot of $(\ln t_*(n))/\s$ against $n^{1/2}$.
The slope of the blue line is the same as that of Figure~\ref{fx}.}
\end{center}
\end{figure}

Figure~\ref{ft} shows a plot of $(\ln t_*(n))/\s$ against $n^{1/2}$,
where the crossover time $t_*(n)$ has been defined by the condition $\rho_n(t)=\rho_n/\e$,
as said above.
This plot is a dynamical analogue of Figure~\ref{fx},
where the system size $N$ is replaced by the distance $n$ traveled by the particle,
whereas time $t$ replaces the inverse of the energy gap $E_1$.
It is clearly far more difficult to get accurate dynamical data
than spectral ones of the same quality.
In spite of this, and although the available data span a rather limited range,
they however point toward an asymptotic linear growth,
whose slope is nearly independent of disorder.
These observations support the prediction~(\ref{nstar}),
again up to a possible weak residual dependence on $\s$.
The slope of the blue line
has been chosen for definiteness to be the same as that of Figure~\ref{fx}.

\begin{figure}[!ht]
\begin{center}
\includegraphics[angle=-90,width=.45\linewidth]{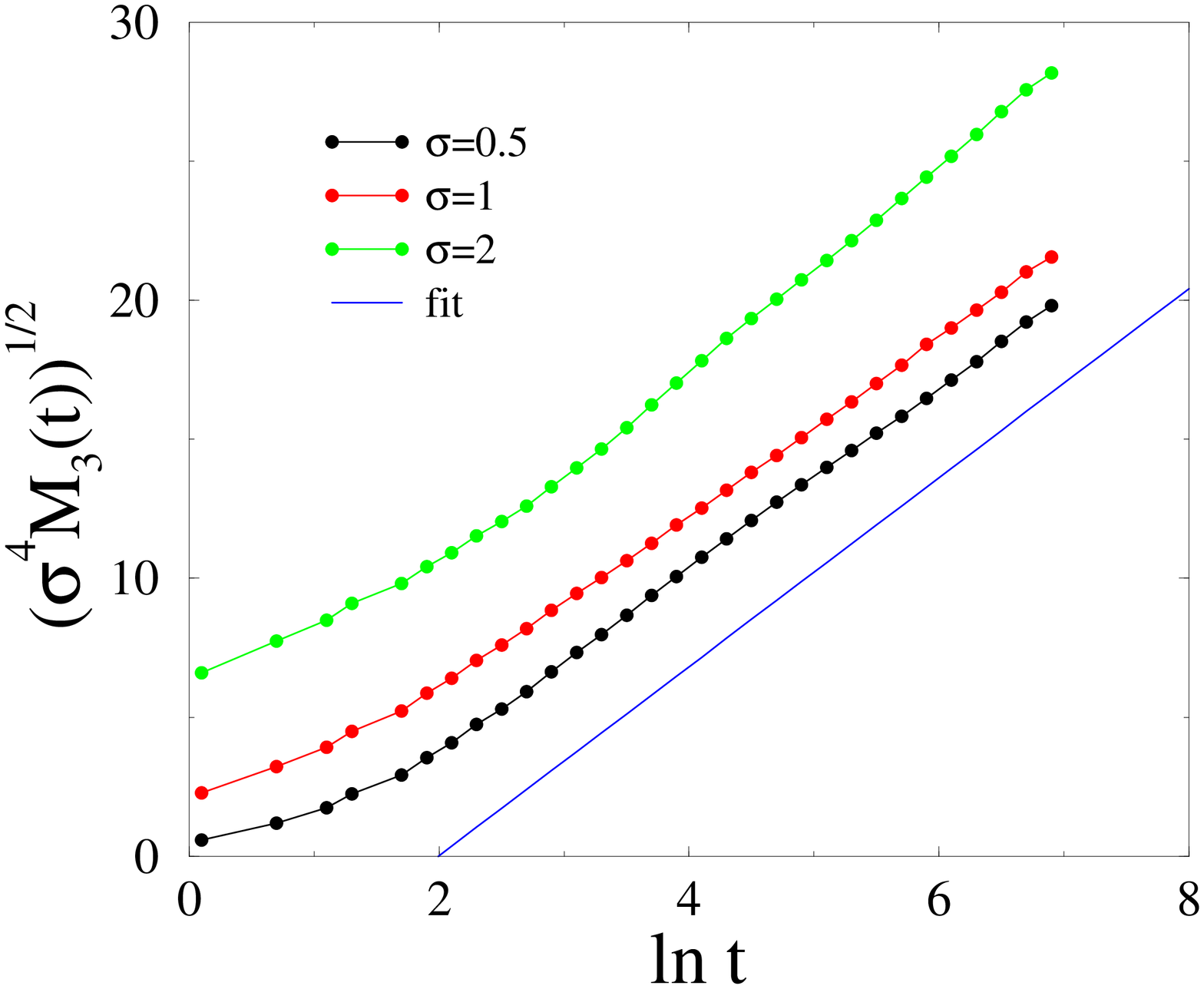}
\qquad
\includegraphics[angle=-90,width=.45\linewidth]{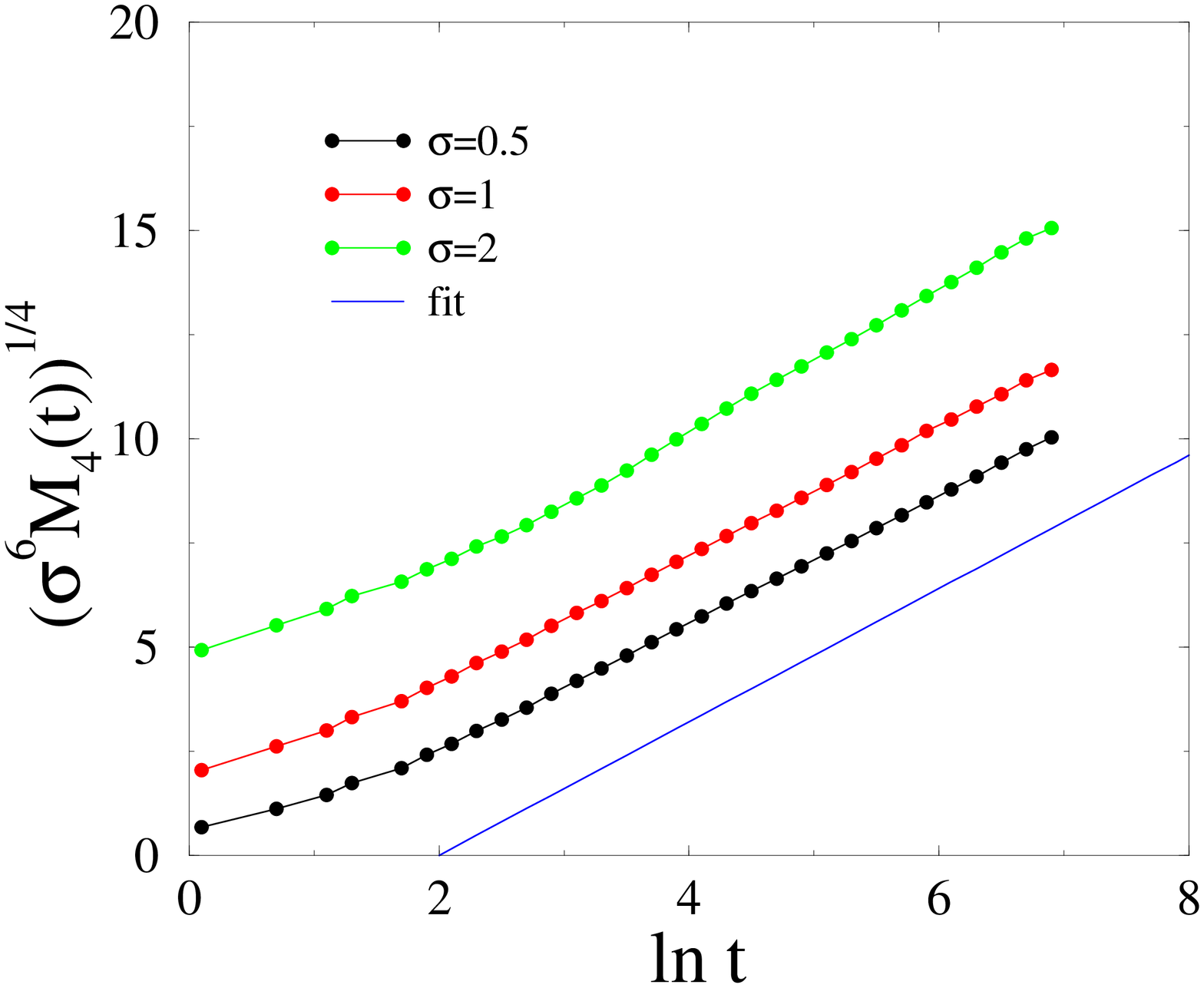}
\caption{\label{fmom34}
Plot of $(\s^4 M_3(t))^{1/2}$ (left) and $(\s^6 M_4(t))^{1/4}$ (right), against $\ln t$.
Straight lines: common least-square fits yielding the amplitudes given in the text.}
\end{center}
\end{figure}

Figure~\ref{fmom34} shows plots of
$(\s^4 M_3(t))^{1/2}$ (left) and $(\s^6 M_4(t))^{1/4}$ (right), against $\ln t$.
The data exhibit linear growth laws,
whose slopes seem independent of $\s$, at least to a good approximation,
thus corroborating~(\ref{M34}).
The slopes of common least-square fits yield $a_3\approx 12$ and $a_4\approx 6$.

\begin{figure}[!ht]
\begin{center}
\includegraphics[angle=-90,width=.45\linewidth]{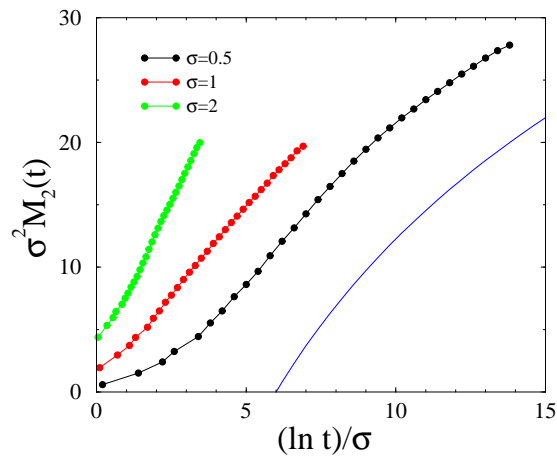}
\caption{\label{fmom2}
Plot of $\s^2 M_2(t)$ against $(\ln t)/\s$.
The blue line illustrates the expected logarithmic growth~(\ref{Mq}).}
\end{center}
\end{figure}

Figure~\ref{fmom2} shows a plot of $\s^2 M_2(t)$ against $(\ln t)/\s$.
This quantity is predicted in~(\ref{Mq}) to grow logarithmically,
with an amplitude independent of the disorder strength~$\s$.
The available data cannot lead to a firm conclusion in this marginal situation.
They however show a trend toward the predicted behavior,
shown by the blue line (up to an unknown multiplicative factor).

The estimate~(\ref{nstar}) of the dynamical crossover scale $n_*(t)$
provides the opportunity of putting the present problem
in perspective with the Sinai walk~\cite{Sin},
namely the diffusion of a classical particle in a one-dimensional
quenched random potential $W(x)$, modeled as a Brownian motion such that
$\ave{W(x)^2}=\sigma^2x$.
The Sinai particle is known to have a very slow anomalous motion,
such that $x_*(t)=\bigl(\ave{\mean{x(t)^2}}\bigr)^{1/2}\sim(\ln t)^2/\s^2$.
This estimate bears a striking resemblance with our crossover scale $n_*(t)$.
Both models are indeed known to share a certain number of common features~\cite{revs}.

The different behaviors of the moments $M_q(t)$ for $q<q_0$ and $q>q_0$
emphasized in~(\ref{Mq}) (with $q_0=2$) can be referred to as {\em bi-fractality}.
This word was already used in the framework of a free quantum particle
described by a tight-binding model in one dimension~\cite{L}.
In this situation,
the moments of the position of the particle exhibit ballistic scaling,
i.e., $M_q\sim t^q$ for all positive $q$.
The moments of the probability density however exhibit a non-trivial scaling:
\begin{equation}
\label{Sq}
S_q(t)=\sum_n\ave{\abs{\psi_n(t)}^{2q}}\sim
\left\{\matrix{
t^{-(q-1)}\hfill&(q<2),\cr
\frad{\ln t^\blanc}{t_\blanc}\hfill&(q=2),\cr
t^{-(2q-1)/3}\hfill\quad&(q>2).
}\right.
\end{equation}
This behavior
is due to the ballistic peaks of the wavefunction $\psi_n(t)$ around $n=\pm2t$
[see~(\ref{psiairy})].
These peaks bring negligible (resp.~dominating) contributions for $q<2$ (resp.~$q>2$).
The word {\em bi-fractal} had already appeared
in the context of the distribution of matter in the Universe~\cite{BS}.

Coming back to our findings,
the main consequence of the divergence of the localization length at the band center
turns out to be the critical nature of the asymptotic density profile of the particle
in the long-time regime,
which falls off as a power of distance,
as $\rho_n\sim1/\abs{n}^{q_0+1}$.
As long as we probe a moment $M_q(t)$ with $q<q_0$,
the fat power-law tail of the latter profile is irrelevant, and so $M_q(t)$ remains finite.
The marginal moment ($q=q_0$) slowly diverges with time,
as the logarithm of the cutoff scale $n_*(t)$.
For $q>q_0$, the divergence of the localization length
is somewhat faster, as $M_q(t)\sim(n_*(t))^{q-q_0}$.

Finally, there is a qualitative difference
between the kind of anomalous diffusion implied
by the bi-fractal behavior met in the present work
[see~(\ref{Mq}),~(\ref{Mqdim}),~(\ref{Mqdim1})]
and conventional anomalous diffusion.
The latter phenomenon is usually
characterized by a single growing length scale~$\ell(t)$,
so that $M_q(t)\sim\ell(t)^q$ for all positive $q$.
In the case of self-similar growth,
like e.g.~in fractional diffusion,
one has $\ell(t)\sim t^\alpha$,
and so $M_q(t)\sim t^{\alpha q}$ for all positive $q$~[see~\cite{PR}
and the references therein].

\section{The random-dimer model}

We now turn to another example of a one-dimensional tight-binding model with
unusual localization properties,
namely the random-dimer model~\cite{PP,B,FH,Sveta}.
The time-dependent Schr\"odinger equation reads
\begin{equation}
\label{RDI}
\ii\,\dot{\psi}_n=\psi_{n+1}+\psi_{n-1}+V_n\psi_n,
\end{equation}
where the site energies $V_n$ follow a sequence of dimers,
i.e., they take random values on {\em pairs} of consecutive sites.
In other words,
\begin{equation}
\label{vdim}
V_{2k}=V_{2k+1}=V\eps_k,
\end{equation}
where $\eps_k$ takes values $+1$ and $-1$ with equal probabilities,
and $V>0$ for definiteness.

\subsection{Spectral properties revisited}

The time-independent problem is defined by the eigenvalue equation
\begin{equation}
\label{RDS}
E\psi_n=\psi_{n+1}+\psi_{n-1}+V_n\psi_n.
\end{equation}
It was shown first by Dunlap {\em et~al}~\cite{PP}
that the random-dimer model exhibits a transparency phenomenon at $E=\pm V$,
and extended states near these special energies for $V\le1$.
This phenomenon is best described in terms of transfer matrices.
The tight-binding equation~(\ref{RDS}) can be recast as
\begin{equation}
\label{TM}
\pmatrix{\psi_{n+1}\cr\psi_n}=T(E,V_n)\pmatrix{\psi_n\cr\psi_{n-1}},
\qquad
T(E,V)=\pmatrix{E-V&-1\cr 1&0\cr}.
\end{equation}

The generic localization properties of the
customary Anderson model with diagonal disorder,
where the site energies $V_n$ are i.i.d.~variables,
can be related to the fact that the transfer matrices associated with single sites
do not commute:
\begin{equation}
[T(E,V_1),T(E,V_2)]=(V_1-V_2)\pmatrix{0&1\cr1&0}.
\end{equation}
The transfer matrices $T(E,V)^2$ and $T(E,-V)^2$ associated with the dimers however obey
\begin{equation}
[T(E,V)^2,T(E,-V)^2]=2V(E^2-V^2)\pmatrix{0&1\cr1&0}.
\end{equation}
This identity underlines that something special occurs
for the dimer model at $E=\pm V$.
Owing to symmetry it is sufficient to consider the case $E=V$.
We have then
\begin{equation}
T(E,V)^2=-\pmatrix{1&0\cr0&1},
\end{equation}
so that adding a dimer with $\eps=+1$ amounts to changing the sign of the wavefunction.
The system therefore supports extended states at (and near) $E=V$,
provided this energy lies in the spectrum of the other species of dimers ($\eps=-1$).
For the sites belonging to the latter dimers,~(\ref{RDS}) reads
\begin{equation}
\label{other}
(E+V)\psi_n=\psi_{n+1}+\psi_{n-1}.
\end{equation}
For $E=V$, the dispersion law of the latter equation reads $V=\cos q$.
The condition for having extended states therefore reads $V\le1$.

The following properties are known for the random-dimer model~\cite{PP,B,FH,Sveta}.

\begin{itemize}

\item
In the {\em generic situation} ($V<1$),
the special energy $E=V$ is inside the band of~(\ref{other}).
The localization length is known to diverge quadratically and symmetrically as
\begin{equation}
\label{gdos}
\xi(E)\sim\frac{1}{(E-V)^2}\qquad(E\to V),
\end{equation}
whereas the density of states is regular at $E=V$.

Neglecting localization effects in the immediate vicinity of the special energy,
the dispersion law of~(\ref{other}) reads $E+V=2\cos q$.
As a consequence, the dynamics of the effectively extended states
is characterized by the ballistic (group) velocity
\begin{equation}
\label{vdef}
v=\bigabs{\frac{\dd E}{\dd q}}_{E=V}=2(1-V^2)^{1/2}.
\end{equation}

\item
In the {\em critical situation} ($V=1$),
the special energy $E=1$ meets the band edge of~(\ref{other}),
and so the velocity $v$ vanishes.
The properties of the spectrum are very asymmetric
in the vicinity of the special energy.
The localization length and the density of states respectively diverge as
\begin{equation}
\label{cdos}
\xi(E)\sim\frac{1}{1-E},\qquad\rho(E)\sim\frac{1}{(1-E)^{1/2}}\qquad(E\to1^-),
\end{equation}
whereas the density of states is exponentially small as $E\to1^+$.

\end{itemize}

Consider now, along the lines of Section~2,
a finite sample made of an even number~$N$ of sites, covered by $N/2$ dimers.
The number of eigenstates which are effectively extended grows as $N^{1/2}$,
both in the generic and in the critical situation~\cite{B}.
In the generic situation ($V<1$), the eigenstates such that the difference
$\delta=E-V$ is smaller than $N^{-1/2}$ have a localization length $\xi$ larger than $N$.
As the density of states is finite at $E=V$,
the number of such effectively extended states grows as $N\delta\sim N^{1/2}$.
Different estimates yield the same outcome in the critical situation ($V=1$).
The effectively extended states are now such that the difference
$\delta=E-1$ is negative and of the order of $1/N$.
As the density of states diverges according to~(\ref{cdos}),
the number of effectively extended states grows as $\delta^{-1/2}\sim N^{1/2}$.

The above property will be the cornerstone of our heuristic analysis
of the dynamical problem.
It is therefore worth checking its range of applicability by means of numerical data.
Let $E^\alpha$ be the $N$ energy eigenvalues of a finite chain
made of an even number $N$ of sites,
and $\psi_n^\alpha$ be the corresponding (real and normalized) eigenvectors.
We define the participation ratio $P^\alpha$ of state number $\alpha$ as
\begin{equation}
\label{ipr}
P^\alpha=\frad{1}{\sum_n(\psi_n^\alpha)^4}.
\end{equation}
This quantity is known to provide a measure of the extension of the eigenstate $\alpha$,
i.e., of the number of sites which participate in the wavefunction $\psi^\alpha$,
in the sense that $\psi_n^\alpha$ takes appreciable values~\cite{ipr}.
The participation ratio scales as $N$ for an extended state,
whereas it is of order unity for a localized state.
As a consequence, the disorder-averaged mean participation ratio
\begin{equation}
\label{iprave}
\ave{P}=\frac1N\sum_\alpha\ave{P^\alpha}
\end{equation}
provides an operational measure of the number
of effectively extended states on a finite sample of size $N$.

\begin{figure}[!ht]
\begin{center}
\includegraphics[angle=-90,width=.45\linewidth]{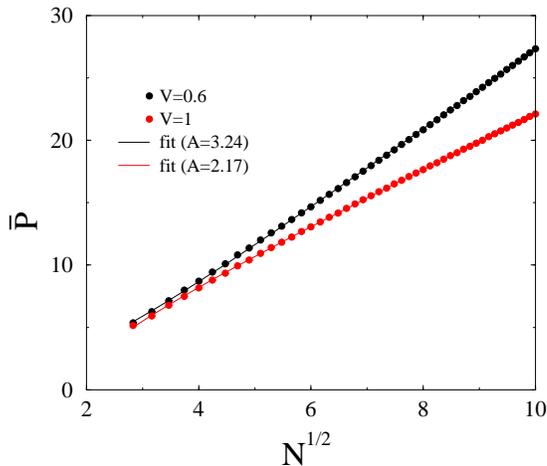}
\caption{\label{fdipr}
Plot of the disorder-averaged mean participation ratio $\ave{P}$
against $N^{1/2}$, for $V=0.6$ (generic situation) and $V=1$ (critical situation).
Curves going almost exactly through the data points:
fits to the form $\ave{P}=AN^{1/2}+B+CN^{-1/2}$.}
\end{center}
\end{figure}

Figure~\ref{fdipr} shows a plot of $\ave{P}$ against $N^{1/2}$.
Data have been obtained by means of a numerical diagonalization of the Hamiltonian matrix
on finite chains with periodic boundary conditions.
The observed asymptotic linear growth confirms that the number
of extended states follows the predicted $N^{1/2}$ growth law,
both in the generic and in the critical situation,
with prefactors of order unity, down to system sizes as small as~10.

\subsection{Dynamical properties}

We now move to the investigation of the quantum dynamics of the random-dimer model.
We again assume that the particle starts at the origin, according to~(\ref{t=0}).

\subsection*{The generic situation ($V<1$)}

The dynamical properties of the random-dimer model in the generic situation
have been the subject of recent rigorous works~\cite{Sveta}.
We proceed along an entirely different route
and employ the following heuristic line of reasoning.

On a large but finite sample of size~$N$,
the model has a number of order $N^{1/2}$ of effectively extended states.
The projection of the initial wavefunction~(\ref{t=0}) onto extended states
therefore scales as the weight of those states in the whole spectrum,
i.e., as $N^{-1/2}$.
This component of the wavefunction spreads ballistically
with velocity $v$ in a transient regime ($t\ll N$).
Now, considering an infinite system but a finite observation time $t$,
we are led to the picture that some random projection
of the initial wavefunction~(\ref{t=0}),
carrying a fraction of order $(vt)^{-1/2}$ of the total intensity,
spreads ballistically.
This picture translates into the following scaling form
for the disorder-averaged density profile:
\begin{equation}
\label{rhosca}
\rho_n(t)\approx\frac{1}{(vt)^{3/2}}\;F\left(\frac{n}{vt}\right)
=\frac{1}{\abs{n}^{3/2}}\;G\left(\frac{n}{vt}\right),
\end{equation}
in the scaling regime where both $n$ and $t$ are large and proportional.

\begin{figure}[!ht]
\begin{center}
\includegraphics[angle=-90,width=.45\linewidth]{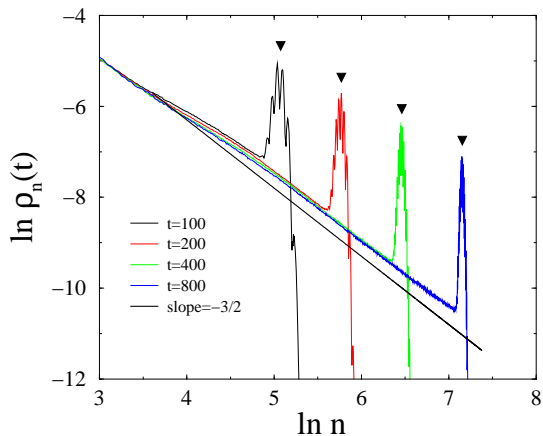}
\caption{\label{fdprofil}
Log-log plot of the disorder-averaged density $\rho_n(t)$ against distance $n$,
for $V=0.6$ and various observation times.
Arrowheads: expected positions of the ballistic peaks ($n=vt$).
Straight line with slope (-3/2): asymptotic power-law profile~(\ref{rhodim}).}
\end{center}
\end{figure}

The asymptotic disorder-averaged density profile $\rho_n$ thus falls off as a power law:
\begin{equation}
\label{rhodim}
\rho_n\sim\frac{1}{\abs{n}^{3/2}}\,.
\end{equation}
Figure~\ref{fdprofil} shows a log-log plot of the disorder-averaged density $\rho_n(t)$
against distance~$n$, for $V=0.6$ (so that $v=1.6$) and times $t=100$, 200, 400, and 800.
The data exhibit sharp ballistic peaks around the expected locations
$n=vt$ (arrowheads).
Both the width and the structure of these peaks are reminiscent
of the behavior~(\ref{psiairy}) of the wavefunction in the absence of disorder.
For $t>n/v$, the data then soon saturate
to the predicted asymptotic power-law profile~(\ref{rhodim}).

The disorder-averaged moments $M_q(t)$ of the position of the particle at time $t$
grow with time for $q\ge1/2$,
whereas they saturate to finite values for $q<1/2$:
\begin{equation}
\label{Mqdim}
M_q(t)\sim
\left\{\matrix{
(vt)^{(2q-1)/2}\hfill\quad&(q>1/2),\cr
\ln(vt)\hfill&(q=1/2),\cr
\hbox{finite}\hfill&(q<1/2).
}\right.
\end{equation}
These predictions, and especially the power-law divergence with exponent $(2q-1)/2$,
are in agreement with recent rigorous results~\cite{Sveta}.
We thus have in particular
\begin{equation}
\label{M12}
M_1(t)\approx b_1 t^{1/2},\qquad
M_2(t)\approx b_2 t^{3/2}.
\end{equation}

\begin{figure}[!ht]
\begin{center}
\includegraphics[angle=-90,width=.45\linewidth]{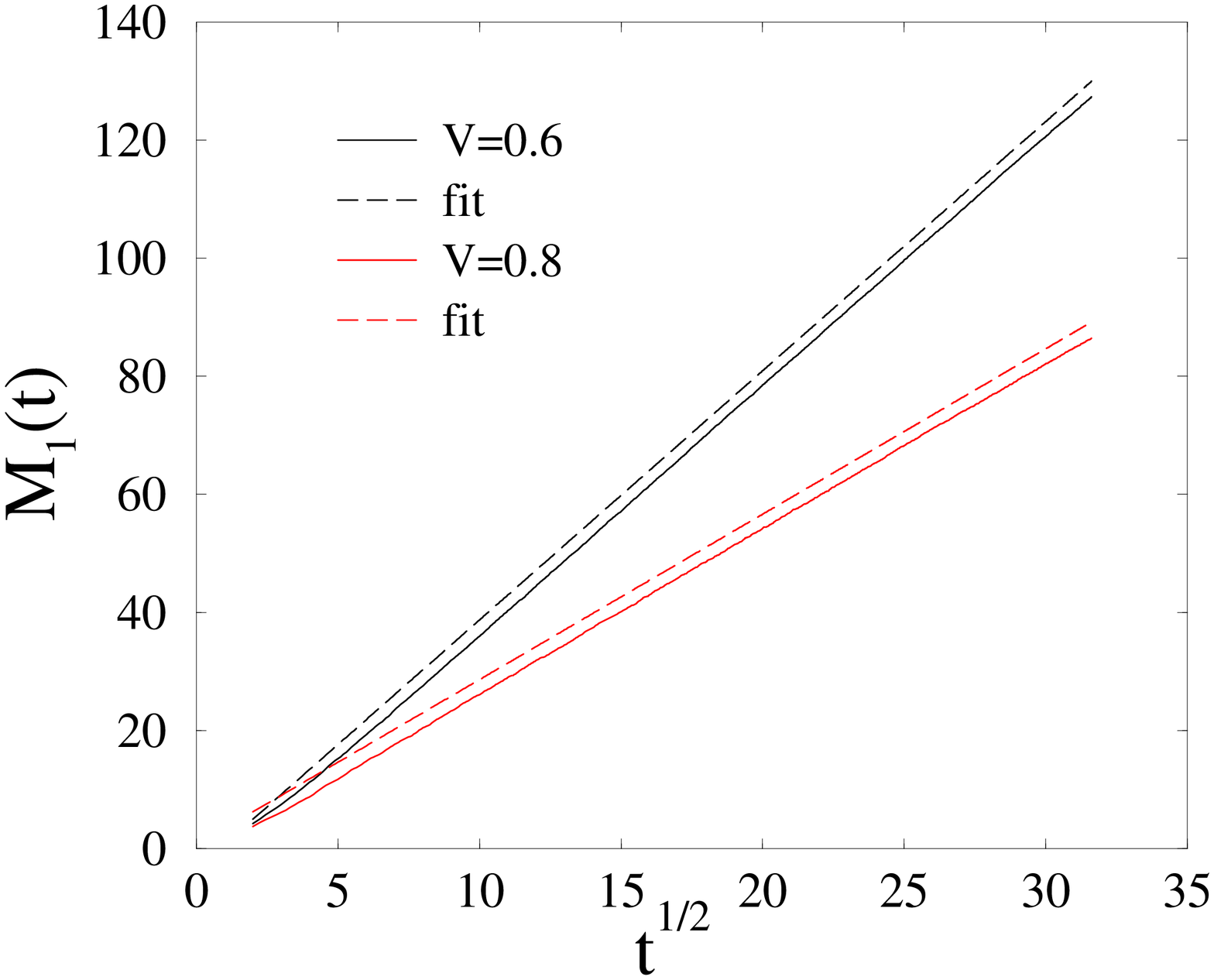}
\qquad
\includegraphics[angle=-90,width=.45\linewidth]{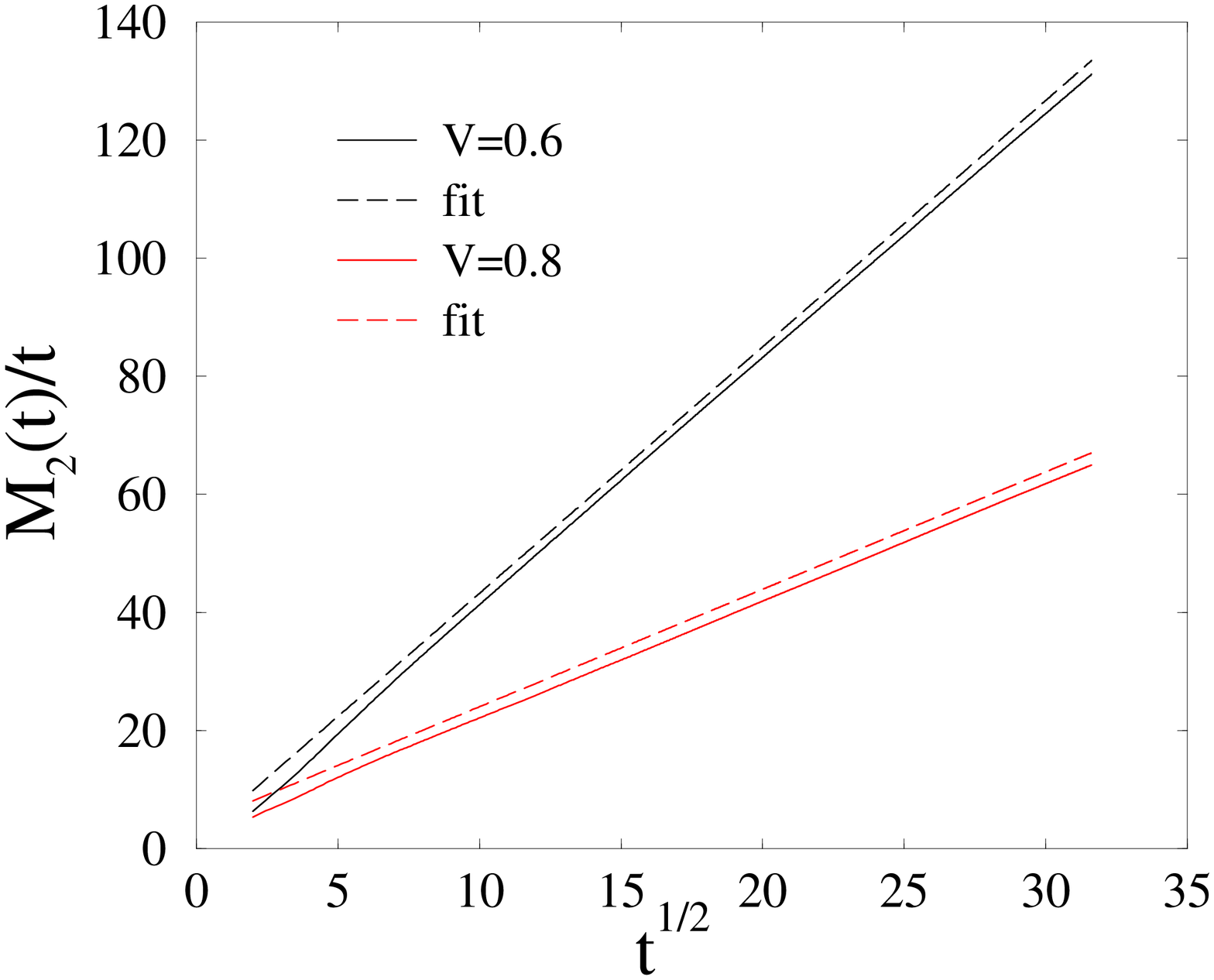}
\caption{\label{fdmom12}
Plots of $M_1(t)$ (left) and of $M_2(t)/t$ (right) against $t^{1/2}$,
for $V=0.6$ and $V=0.8$.
Straight lines (slightly translated for readability):
least-square fits yielding the amplitudes given in the text.}
\end{center}
\end{figure}

Figure~\ref{fdmom12} shows plots of $M_1(t)$ (left) and $M_2(t)/t$ (right)
against $t^{1/2}$, for $V=0.6$ (so that $v=1.6$) and $V=0.8$ (so that $v=1.2$).
The plotted data exhibit linear growth laws with a very high accuracy.
The corresponding amplitudes read $b_1\approx4.2$ and $b_2\approx4.2$ for $V=0.6$,
and $b_1\approx2.8$ and $b_2\approx2.0$ for $V=0.8$.

\subsection*{The critical situation ($V=1$)}

Consider now the critical situation ($V=1$), where the velocity $v$ vanishes.
It is worth mentioning that no rigorous result is available in this case.

The previous line of reasoning can be adapted in the following heuristic way.
On a large but finite sample of size $N$,
the model has a number of order $N^{1/2}$ of effectively extended states.
The projection of the initial wavefunction~(\ref{t=0}) onto those states
therefore again scales as $N^{-1/2}$.
The latter component of the wavefunction is again expected to spread ballistically
in a transient regime, with a non-zero effective velocity~$v_*(N)$
due to the finiteness of the system.
The latter can be estimated as follows.
The effectively extended states have an energy $E$ such that $1-E\sim1/N$,
whereas the velocity $v$ is such that $1-V\approx v^2/8$ [see~(\ref{vdef})].
Equating both estimates of the energy shift $1-E\sim1-V$ yields $v_*(N)\sim N^{-1/2}$,
implying therefore the existence of a crossover time $t_*(N)\sim N/v_*(N)\sim N^{3/2}$.
The growth law of this time scale is intermediate between the ballistic one ($t\sim N$)
and the diffusive one ($t\sim N^2$).

\begin{figure}[!ht]
\begin{center}
\includegraphics[angle=-90,width=.45\linewidth]{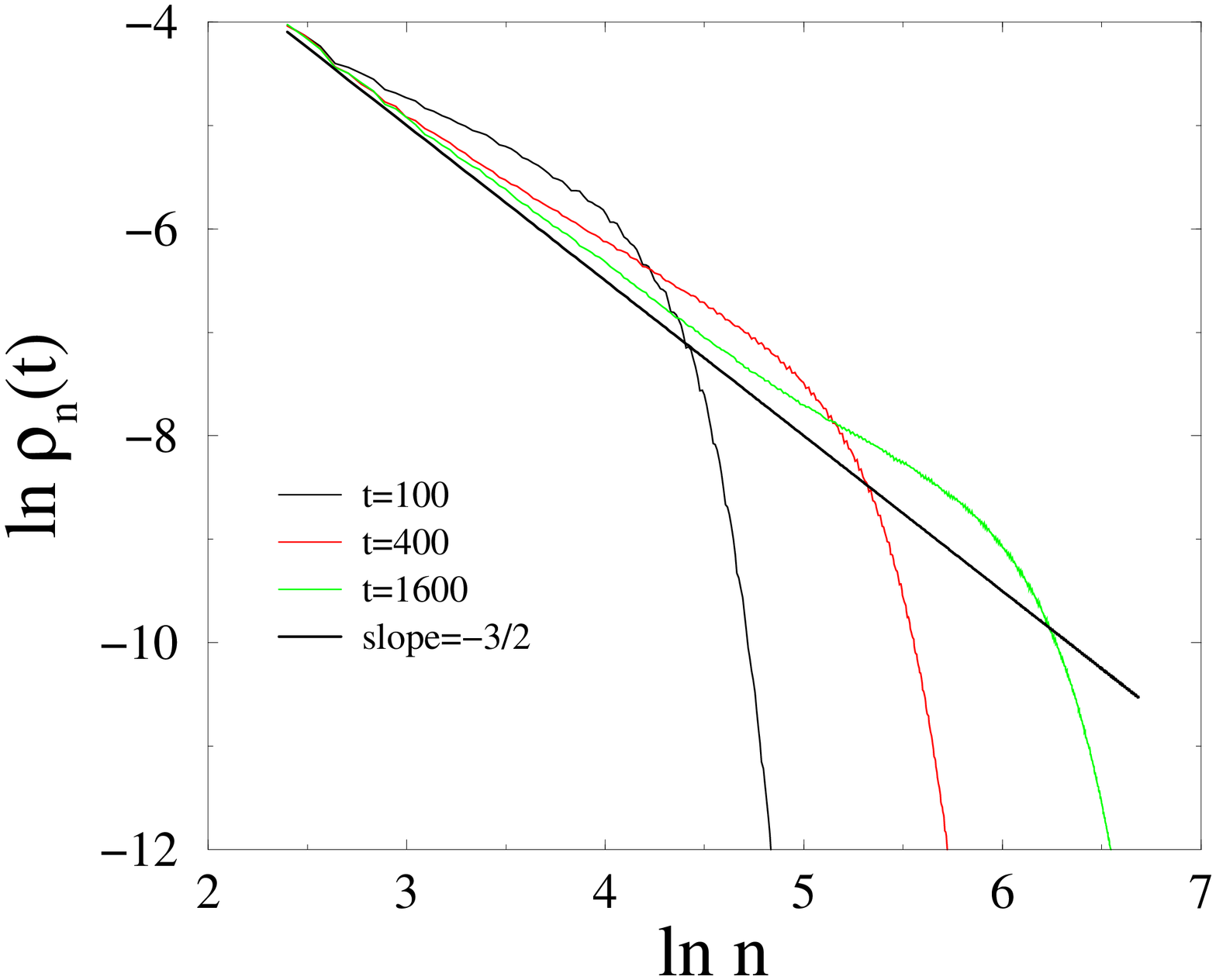}
\qquad
\includegraphics[angle=-90,width=.45\linewidth]{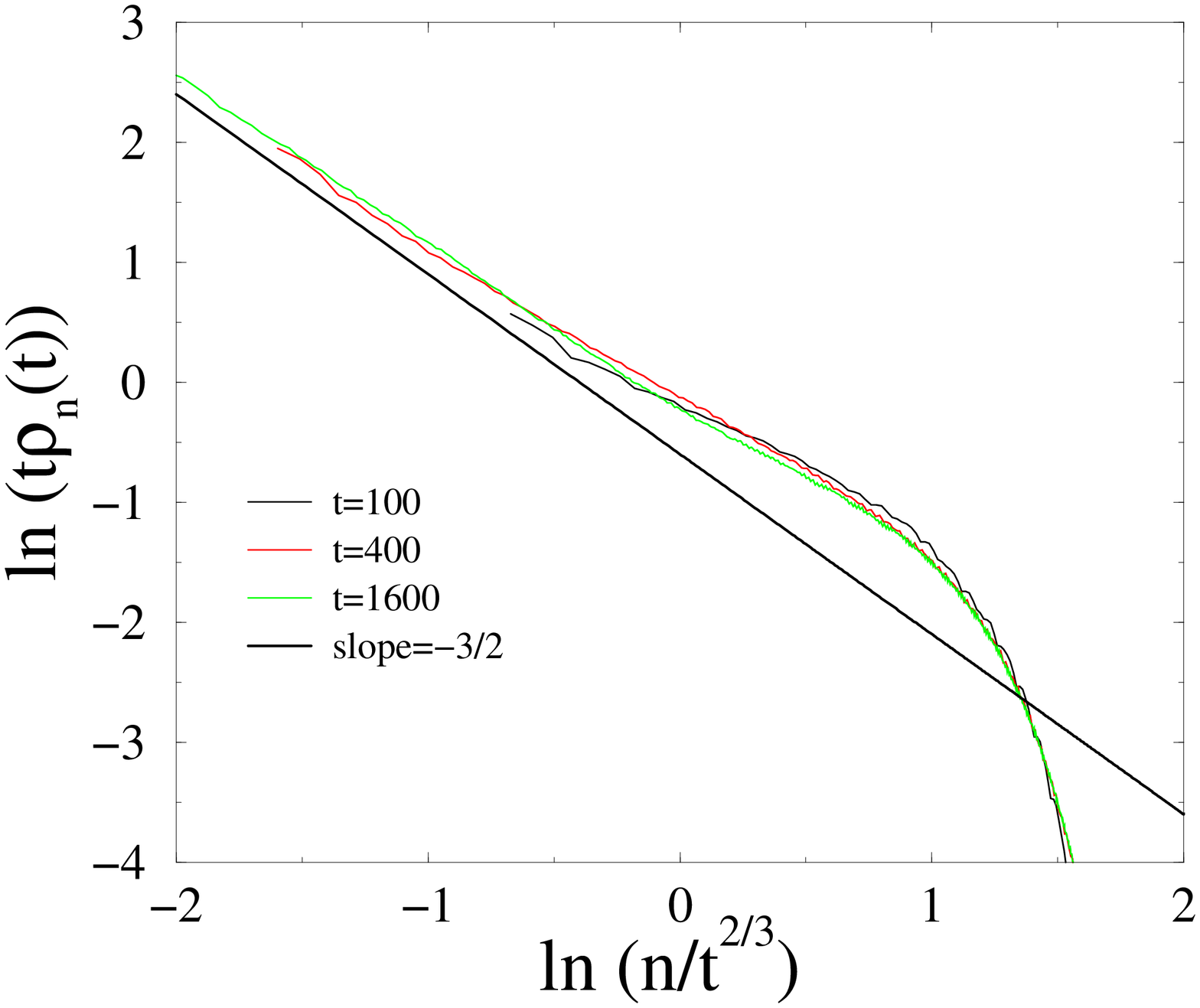}
\caption{\label{fd1profil}
Left: Log-log plot of the disorder-averaged density $\rho_n(t)$ against $n$,
for $V=1$ and various observation times.
Straight line with slope (-3/2): asymptotic power-law profile~(\ref{rhodim}).
Right: Rescaled plot of the same data demonstrating the scaling law~(\ref{rhosca1}).}
\end{center}
\end{figure}

The above argument suggests the following scaling form
for the disorder-averaged density profile:
\begin{equation}
\label{rhosca1}
\rho_n(t)\approx\frac{1}{t}\;F_1\left(\frac{n}{t^{2/3}}\right)
=\frac{1}{\abs{n}^{3/2}}\;G_1\left(\frac{n}{t^{2/3}}\right).
\end{equation}
The asymptotic disorder-averaged density profile $\rho_n$
again falls off according to~(\ref{rhodim}).
The left panel of Figure~\ref{fd1profil}
shows a log-log plot of the disorder-averaged density $\rho_n(t)$
against distance $n$, for times $t=100$, 400, and 1600.
The data converge toward the predicted asymptotic power-law profile~(\ref{rhodim}).
A comparison with Figure~\ref{fdprofil} however shows
that the sharp ballistic peaks are replaced by broad shoulders.
The right panel of Figure~\ref{fd1profil} shows a rescaled plot of the same data.
The very good quality of the observed data collapse,
especially for the larger values of the ratio $n/t^{2/3}$,
strongly supports the scaling law~(\ref{rhosca1})
derived by heuristic means.

\begin{figure}[!ht]
\begin{center}
\includegraphics[angle=-90,width=.45\linewidth]{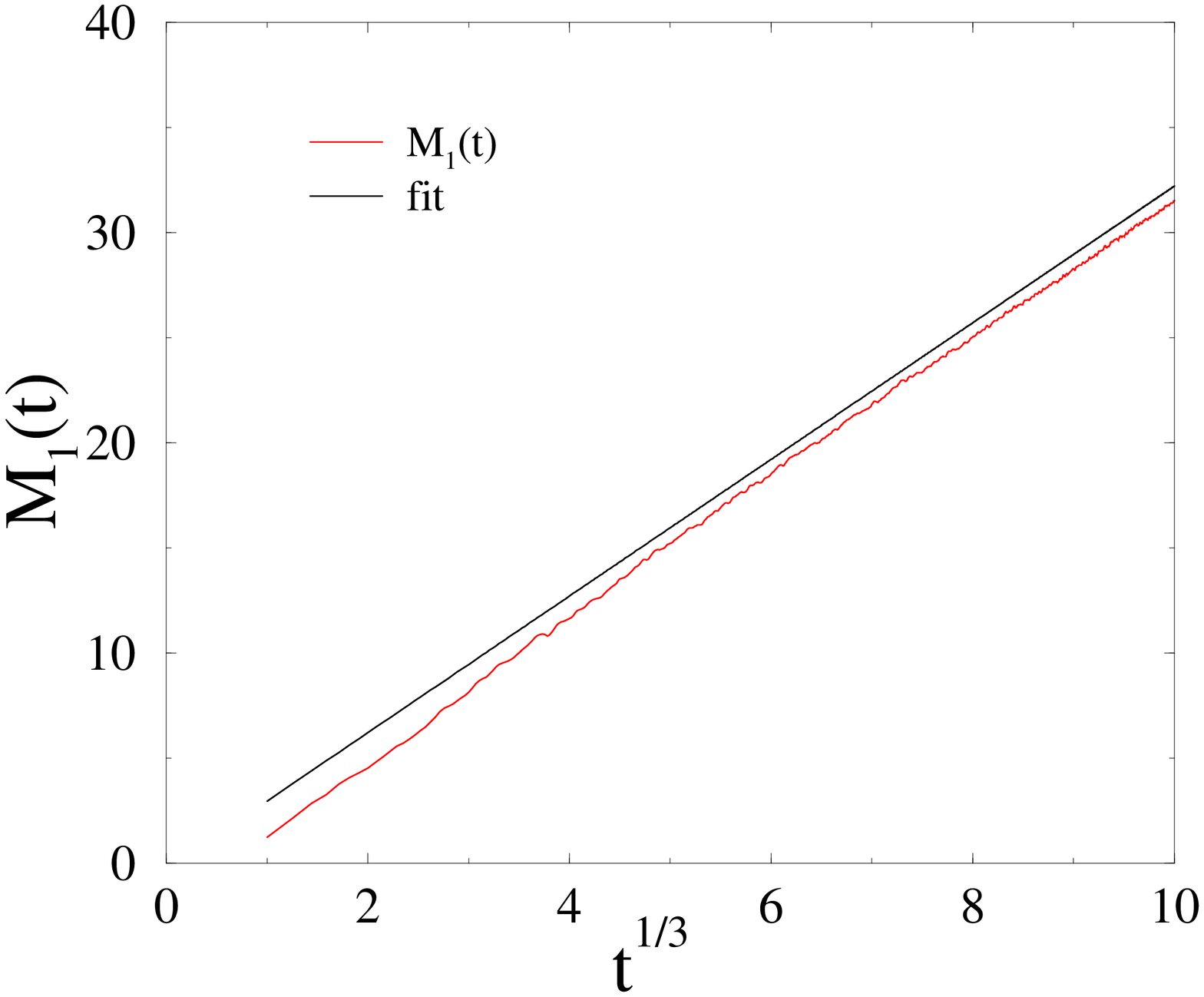}
\qquad
\includegraphics[angle=-90,width=.45\linewidth]{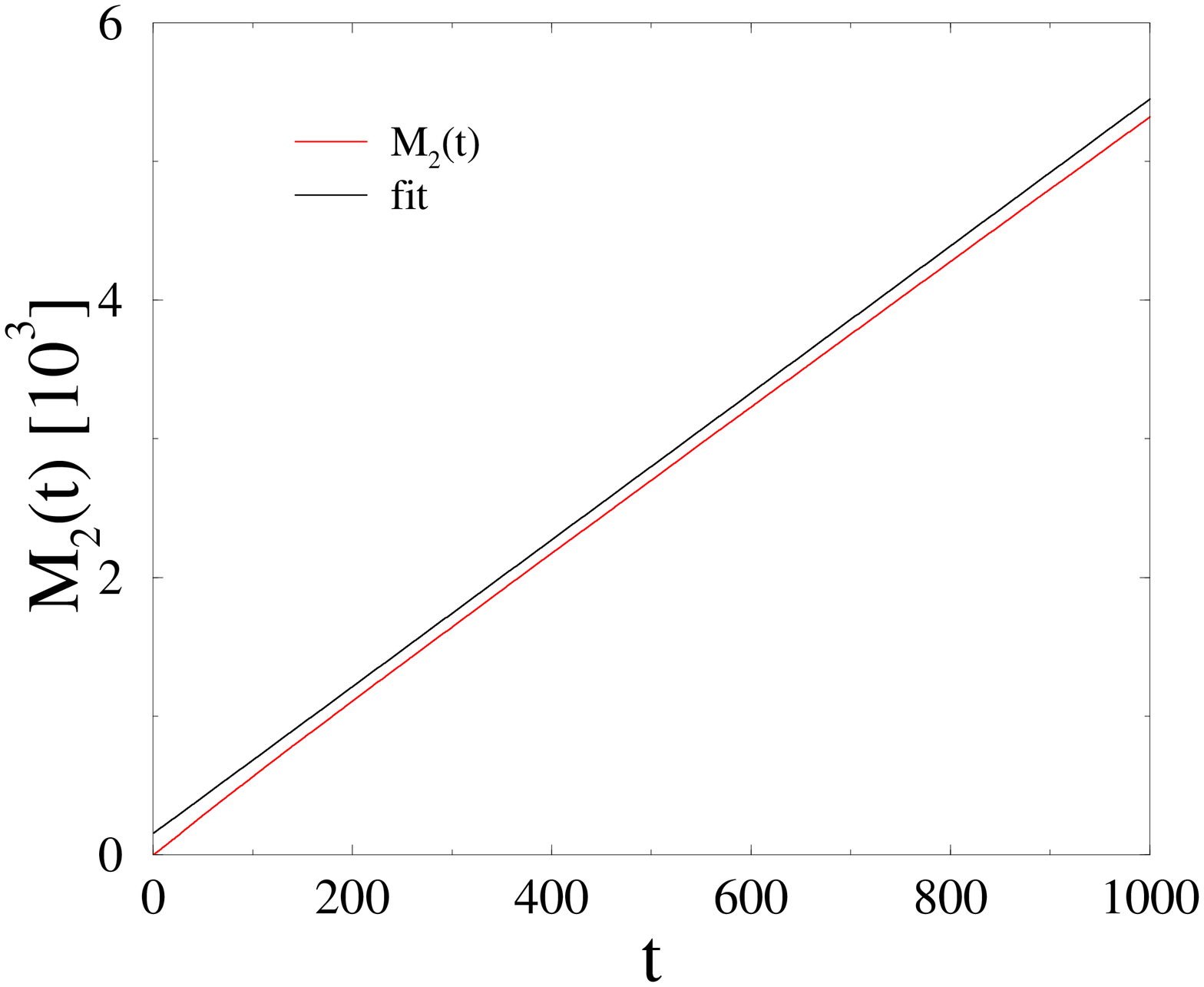}
\caption{\label{fd1mom}
Plots of $M_1(t)$ against $t^{1/3}$ (left) and of $M_2(t)$ against $t$ (right)
for $V=1$.
Straight lines (slightly translated for readability):
least-square fits yielding the amplitudes given in the text.}
\end{center}
\end{figure}

As a consequence of~(\ref{rhosca1}),
the disorder-averaged moments $M_q(t)$ of the position of the particle at time $t$
grow with time for $q\ge1/2$,
whereas they saturate to finite values for $q<1/2$:
\begin{equation}
\label{Mqdim1}
M_q(t)\sim
\left\{\matrix{
t^{(2q-1)/3}\hfill\quad&(q>1/2),\cr
\ln t\hfill&(q=1/2),\cr
\hbox{finite}\hfill&(q<1/2).
}\right.
\end{equation}
We have in particular
\begin{equation}
\label{M121}
M_1(t)\approx c_1 t^{1/3},\qquad
M_2(t)\approx c_2 t.
\end{equation}
Figure~\ref{fd1mom} shows plots of $M_1(t)$ against $t^{1/3}$ (left)
and $M_2(t)$ against $t$ (right).
The plotted data clearly exhibit linear growth laws,
thus corroborating~(\ref{M121}).
The corresponding amplitudes read $c_1\approx3.2$ and $c_2\approx5.3$.

\section{Discussion}

We have investigated two examples
of disordered one-dimensional Hamiltonians which are exceptions to the common wisdom
on Anderson localization,
in the sense that the localization length diverges at some special energies.

The main goal of the present study has been to provide a unified picture
for the dynamical consequences of this divergence.
The two examples we have considered are very different in nature.
On the chain with off-diagonal disorder (Section~2),
the localization length has a mild logarithmic divergence as $E\to0$.
As a consequence, on a finite sample of $N$ sites,
the most extended eigenstate has an effective localization length
of the order of $N^{1/2}/\s$.
The system therefore supports no effectively extended state.
On the contrary, in the random-dimer model (Section~3),
the localization length has a strong power-law divergence as $E\to\pm V$.
A large but finite sample now supports a large number
of effectively extended states, growing as $N^{1/2}$.
This difference probably explains why the study of the random-dimer model
is far easier than that of the chain with off-diagonal disorder,
both from the mathematical and the numerical standpoints.

The main features of the quantum dynamics put forward in this work
are nevertheless very similar in both models:

\begin{itemize}

\item
The disorder-averaged densities $\rho_n=\ave{\abs{\psi_n(t)}^2}$
converge to non-trivial asymptotic values~$\rho_n$,
which fall off at large distances as a power law of the form
\begin{equation}
\label{rhodisc}
\rho_n\sim\frac{1}{\abs{n}^{q_0+1}}
\end{equation}
[see~(\ref{rho}),~(\ref{rhodim})].

\item
For a large but finite observation time, the power-law density profile is cutoff
at some spatial scale $n_*(t)$ which keeps growing with time.

\item
The disorder-averaged moments of the position of the particle keep growing with time as
\begin{equation}
\label{Mqdisc}
M_q(t)\sim
\left\{\matrix{
(n_*(t))^{q-q_0}\hfill\quad&(q>q_0),\cr
\ln n_*(t)\hfill&(q=q_0),
}\right.
\end{equation}
while they reach finite limits for $q<q_0$
[see~(\ref{Mq}),~(\ref{Mqdim}),~(\ref{Mqdim1})].
This kind of dichotomous behavior can be referred to as {\em bi-fractality}.

\end{itemize}

\begin{table}[!ht]
\caption{Summary of main results on the spectral and dynamical properties
of a quantum particle
on the chain with off-diagonal disorder and on the random-dimer model,
both in the generic situation ($V<1$) and in the critical one ($V=1$).}
\label{summary}
\begin{center}
\begin{tabular}{|l|c|c|c|}
\hline
Model&off-diagonal disorder&\deuxlignes{generic dimers}{($V<1$)}&\deuxlignes{critical dimers}{($V=1$)}\\
\hline
Special energies $E_c$&0&$\pm V$&$\pm1$\\
Localization length $\xi(E)$&$-\ln\abs{E}/\s^2$&$1/(E\mp V)^2$&$1/\abs{E\mp1}$\\
\hline
Asymptotic profile $\rho_n$\hfill&$1/n^3$&$1/n^{3/2}$&$1/n^{3/2}$\\
Index $q_0$\hfill&2&1/2&1/2\\
Crossover scale $n_*(t)$\hfill&$(\ln t)^2/\s^2$&$vt$&$t^{2/3}$\\
Moment $M_q(t)$\ \ ($q>q_0$)\hfill&$(\ln t)^{2(q-2)}/\s^{2(q-1)}$&$(vt)^{(2q-1)/2}$&$t^{(2q-1)/3}$\\
\hline
\end{tabular}
\end{center}
\end{table}

Table~\ref{summary} summarizes our quantitative findings
concerning the chain with off-diagonal disorder and the random-dimer model.

On higher-dimensional bipartite lattices,
tight-binding models with off-diagonal disorder
also exhibit specific features near their band center,
again because of particle-hole symmetry~\cite{G,ITA,H,M}.
In two dimensions, it has been argued that the energy gap~$E_1$
on a finite sample of linear size $L$ scales as $X_1=-\ln E_1\sim(\ln L)^x$.
As a consequence,
the integrated density of states exhibits an essential singularity of the~form
\begin{equation}
\label{dos}
H(E)-1/2\sim\exp(-c(-\ln\abs{E})^{1/x})
\end{equation}
as $E\to0$, i.e., near the band center.
The dynamical exponent has been first predicted by Gade~\cite{G} to be $x=2$,
while more recent studies~\cite{H,M} predict that the exact value of the dynamical
exponent is somewhat smaller: $x=3/2$.
The dimensional scaling Ansatz $H-1/2\sim1/\xi^2$
suggests that the localization length diverges as
$\xi(E)\sim\exp((c/2)(-\ln\abs{E})^{1/x})$.
These rough results fit within the same framework as our one-dimensional findings,
with an asymptotic density profile falling off as
\begin{equation}
\label{rho2}
\rho_n\sim\frac{1}{\s^2\,\abs{\n}^4}\,,
\end{equation}
a critical index $q_0=2$, and bi-fractal moments scaling as
\begin{equation}
\label{Mq2d}
M_q(t)\sim
\left\{\matrix{
\exp((q-2)(c/2)(\ln t)^{2/3})\hfill\quad&(q>2),\cr
(\ln t)^{2/3}\hfill&(q=2),\cr
\hbox{finite}\hfill&(q<2).
}\right.
\end{equation}
One of the chief challenges for future work would be to establish
by means of a more controlled approach the criticality of the stationary density profile
and the validity of~(\ref{Mq2d}) in the two-dimensional situation.

\ackn

We are thankful to Pierre Pujol for valuable discussions.
The work of PLK has been supported by NSF grant CCF-0829541.

\section*{Note added}

After completion of this article we became aware of the recent work~\cite{lepri}
by Lepri {\em et~al}.
These authors investigate the spreading of a wavepacket
in disordered harmonic and anharmonic chains.
In the harmonic case they predict a power-law fall-off
of the energy profile averaged over time and disorder,
with a universal exponent equal to 5/2 for an initial displacement excitation,
and to 3/2 for an initial momentum excitation.
The anomalous spreading and the existence of a non-trivial limiting profile
are demonstrated to be due to the divergence
of the localization length of the phonon modes at low frequency.
The work by Lepri {\em et~al} and the present one
therefore strongly corroborate each other,
as they put forward essentially the same scenario.

We thank Stefano Lepri for having made us aware of this very interesting piece of work.

\end{document}